# A LEVEL TRAINING SET WITH BOTH A COMPUTER-BASED CONTROL AND A COMPACT CONTROLLER

Hayati Mamur[1], Ismail Atacak[2], Fatih Korkmaz[3] and M. R. Amin Bhuiyan[1]

[1]Department of Electrical and Electronics Engineering, Faculty of Engineering,
Manisa Celal Bayar University, 45100, Manisa, Turkey
hayati.mamur@cbu.edu.tr

[2]Department of Computer Engineering, Faculty of Technology,
Gazi University, 06100, Teknikokullar, Ankara, Turkey

[3]Department of Electrical and Electronics Engineering, Faculty of Engineering,
Cankiri Karatekin University, 18100, Cankiri, Turkey

## ABSTRACT

*In engineering education, the combination with theoretical education and practical education is an essential problem. The taught knowledge can be quickly forgotten without an experimental application. In addition, the theoretical knowledge's cannot be easily associated applications by students when they start working in industry. To eliminate these problems, a number of education tools have been developed in engineering education. This article presents a modelling, simulation and practice study of a newly designed liquid level training set developed for the control engineering students to simulate, examine and analyze theoretically and experimentally the controllers widely used in the control of many industrial processes. The newly designed training set combines two control structures, which are a computer-based control and a digital signal processing-based control. The set displays the results related to experiments in real time as well as. These features have made it a suitable laboratory component on which the students can both simulate and test the performance of liquid level control systems by using theoretical different control structures.*

## 1. INTRODUCTION

Today, control systems with a wide usage area ranging from mathematics to processes have become more important with increased industrialization [1], [2]. In industrial processes, many physical variables such as pressure, temperature, level and flow need to be automatically controlled in order to keep their values at a desired level [3], [4]. In this context, automatic control systems are engineering courses that are being taught in different departments of faculties to teach the basic concepts and theorems of control systems [5].

In general, automatic control systems are grouped in two categories: open loop control system and closed loop system [6], [7]. The most important element of the control system is a controller. Personal computers (PCs) connected to data acquisition (DAQ) cards [8]–[11], compact proportional-integral-derivative (PID) control devices [12], programmable logic controllers (PLCs) [13], [14], microcontrollers (μCs) [6], [15], [16] and digital signal processors (DSPs) [17] are the most commonly used controllers in practically all industrial control applications. These controllers interconnect all parts of the physical and non-physical of a system.

Setting of controller parameters, according to a process, has a vital importance in terms of all system performance [18, 19]. Various techniques are employed for setting of these parameters [20]–[22]. Each process needs to be specifically set the controller parameters, but this may not always provide the desired performance on process [23]. Therefore, the setting of controller parameters and the understanding how the parameters have an effect on a process are important for engineering students.

Practicability of control methods and parameter settings through some simulation software and their remote controllability provides convenience to engineering students in the learning of automatic control systems [9], [24]. Moreover, the simulation software's eliminate the need of an experiment set required for engineering courses [25]. However, this type teaching and learning makes difficult the adaptation of engineering students to actual systems. The designating and executing of control methods, controllers and parameters and the observing of the effects to the performance on an actual system in laboratory environment ensure to be more permanent of the learned knowledge for engineering students [26].

For this purpose, Vahidi et al. [25] conducted a study on power transformers by using MATLAB/Simulink environment for incoming engineers in industry and graduate students. Tekin et al. [27] established an internet-based laboratory environment and then performed an induction motor application controlled by a DSP. Stefanovic et al. [11] setup a distance learning laboratory based on LabVIEW for control engineering education. Khairurrijal, et al. [16] developed a temperature control learning set with proportional controller for the education of undergraduate students. They received highly positive feedbacks from the students in the survey conducted on the temperature control learning set. Kayisli et al. [7] developed an education tool with a proportional-integral controller by using MATLAB/Simulink environment for DC-DC converter circuits. Ekici [10] conducted a study on the computer-assisted training method for power analysis courses. Dandil [17] realized a simulation and application study on an induction motor control for undergraduate students. A study conducted by Shyr [26] revealed that the project-based mechatronics learning is more effective in the learning of undergraduate students.

In this study, a liquid level control training set was designed and implemented for control engineering courses. Thanks to the training set, students can both examine the differences between the computer-based control structure with a visual interface and the DSP-based control structure and observe the effects on process performance of the changes in the controller parameters. Thus, two control devices were combined by the training set. Also, both theoretical education and practical education were applied on the training set.

## 2. OBTAINING OF MATHEMATICAL MODEL OF THE TRAINING SET

Before proceeding to the implementation stage, the knowing of how to exhibit a dynamic behaviour of liquid level control systems is vital in terms of the determining of a number of parameters related to the system in the design stage. The dynamic behaviour of a system can be observed and analyzed through some simulation studies after the obtaining of the mathematical model to the system [28]. As in many physical systems, liquid level control systems also exhibit a non-linear dynamic behaviour due to the inherent characteristics [29]. Nevertheless, the theories and theorems in control systems, the most of which can be only applied to linear systems, necessitates to be modelled nonlinear systems as a linear system. When considering small changes in the level control system, the system can be modelled with linear differential equations [30]–[32]. The schematic diagram of the first order liquid level control system controlled is shown in **Fig. 1**.

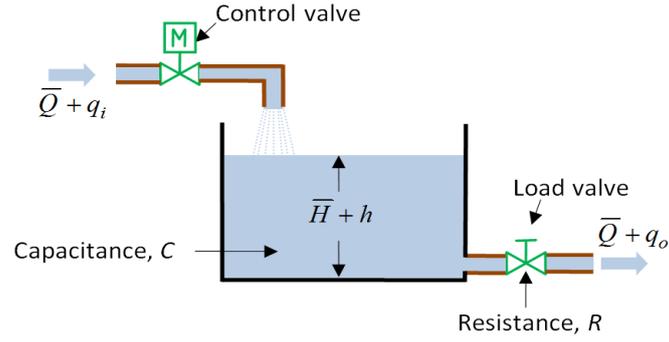

**Figure 1.** A schematic diagram of the liquid level control system

This system consists of a tank with an inlet (control) valve and outlet (load) vane and represents a single input single output (SISO) control system [5]. In the system, while the outflow liquid from the tank is manually controlled through the load vane, the inflow liquid into the tank is adjusted by a proportional valve. Normally, the outflow liquid from the tank is a load which is needed by process and continuously changes due to reasons beyond control. Therefore, the inflow liquid into the tank represents a manipulated variable (MV) depending on the liquid level. The outflow liquid from the tank refers to a load or a disturbance [32]. The top of the liquid tank is open and it has a cylindrical structure. The dimensions and some calculations of the liquid level control tank are given in **Table 1**.

**Table 1.** The dimensions of the tank.

| Properties | Values |
| --- | --- |
| The height of the tank, $h$ (m) | 1 |
| The diameter of the tank, $d$ (m) | 0.15 |
| The cross sectional area of the tank, $A$ (m$^2$) | 0.5063 |
| The capacitance of the tank, $C$ (m$^3$/m) | 0.5063 |
| The volume of the tank, $V$ (m$^3$) | 0.0176 |
| The maximum liquid flow, $Q_o$ (l/s) | 0.5 |
| The maximum liquid flow, $Q_o$ (m$^3$/s) | 0.0005 |
| Resistance, $R$ (s/m$^2$) | 2000 |

The liquid level control system has been modeled taking into account the change in the liquid level, which results from the difference between the inlet flow rate and outlet flow rate of the liquid in the tank. This system can be considered as a simple circuit including a capacity (C) and a resistance (R). By supposing that all initial conditions of the system are equal to zero, the obtained Laplace transform is given by:

$$RCsH(s) + H(s) = RQ_i(s). \qquad (1)$$

If it is considered that height (h) and inflow (qi) are the output and input of the system, respectively, the transfer function is expressed as follows:

$$\frac{H(s)}{Q_i(s)} = \frac{R}{sRC + 1}, \qquad (2)$$

where, RC (τ) is the time constant of the system (s). If outflow (q0) is taken as the output and the input is the same (qi), then the transfer function of the system is defined by:

$$\frac{Q_o(s)}{Q_i(s)} = \frac{1}{RCs+1}. \quad (3)$$

The characteristic equation of the obtained transfer function, simply the denominator of this function, is first order. Therefore, the dynamic behavior of the system is defined in form of time constant. When the calculated R and C values in **Table 1** is substituted in (3), the transfer function of the liquid level control system is achieved by:

$$G(s) = \frac{Q_o(s)}{Q_i(s)} = \frac{1}{1012.6s+1}. \quad (4)$$

The MATLAB/Simulink model of the whole system, which consists of the mathematical model derived for the liquid level control system and the PID controller with the tuned gain parameters, is given in **Fig. 2a**. In **Fig. 2b**, the step response of the system is shown.

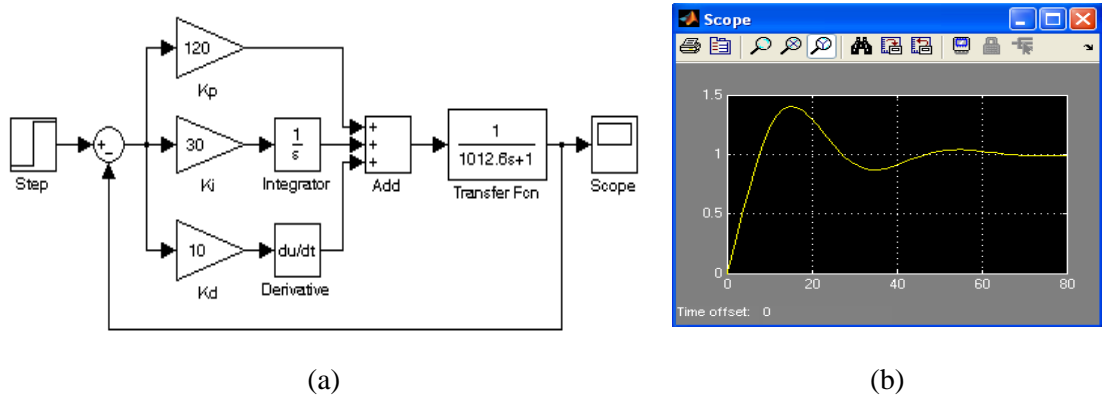

(a)          (b)

**Figure 2**. (a) MATLAB-Simulink model of the liquid level control system with PID controller and (b) Step response of the system

## 3. EXPERIMENTAL SETUP

In **Fig. 3**, the experimental setup and block diagram of the liquid level training set are presented. The equipment list also is given in **Table 2**.

In the system, a NC200 process control device is used as a compact PID controller that is developed for educational purposes by Endcon Company. This device has the setup properties of the fundamental control methods, including on-off control, P control, PD control, PI control and PID control. Also, it has an analog input (AI) of 4-20 mA and an analog output (AO) of 10 V DC. While the AI of the device is connected to a Foxboro differential pressure transmitter which measures the process variable, its AO is applied to a Siemens Acvatix SQS65 proportional vane which controls the process variable. The control signals produced by the control methods in the device are sent to the proportional vane through the AO channel.

Another controller used in control of the liquid level system is a computer-based SCADA control system with DAQ card. In this type control, the level variable in the system is measured by the Foxboro differential pressure transmitter as in the system controlled by the compact PID controller, and then the output of the transmitter is applied to the AI channel of ATS A-4011L DAQ card. The control data that comes from the SCADA on the PC is sent to the Siemens Acvatix SQS65 proportional vane by the AO channel of ATS A-4021L DAQ card. The communication between the DAQ cards and the SCADA software on the PC is done by ATS A-4520L DAQ RS232-RS484 converter The RS232 communication protocol is employed in the data communication from the PC to the DAQ cards. Besides, the RS485 communication protocol is used to transfer the data from the DAQ cards to the PC. The computer-based PID control algorithm is carried out over the DAQ SCADA program developed by the authors.

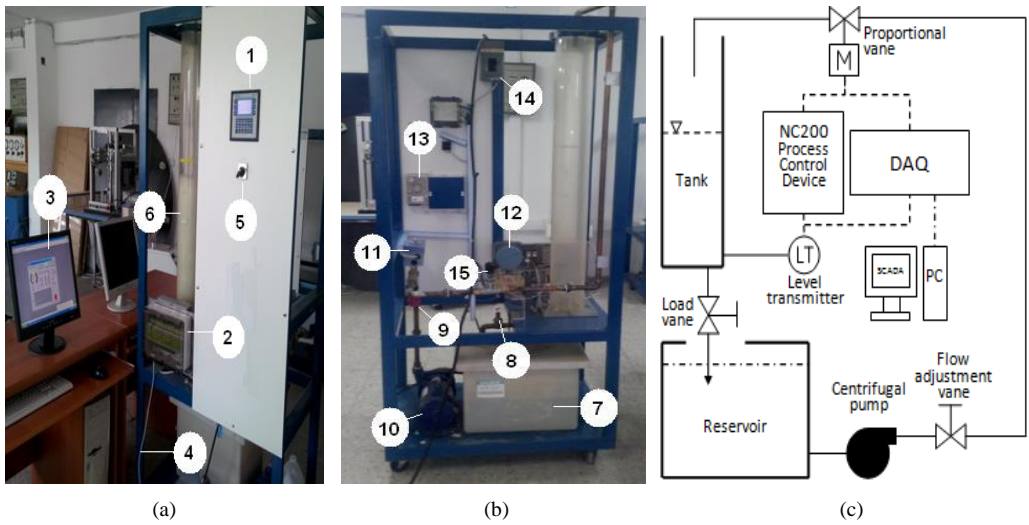

(a)      (b)      (c)

**Figure 3**. The liquid level control training set (a) the front view, (b) the back view and (c) the block diagram

**Table 2.** The equipment list and their properties

| Part no | Name | Properties |
|---|---|---|
| 1 | Compact PID device | ENDCON NC200 |
| 2 | DAQ | ATS A-4520L, 4011L, 4080, 4021L |
| 3 | SCADA | Genie |
| 4 | RS232 communication | 5 m cable |
| 5 | On/off power switch | 10 A |
| 6 | Tank | Fiber glass, $R = 15$ cm, $h = 125$ cm |
| 7 | Reservoir | 40×30×30 cm |
| 8 | Load vane | Manuel |
| 9 | Flow adjustment vane | Manuel |
| 10 | Centrifugal pump | General Electric, 1~ 220 V AC, ½ HP |
| 11 | Proportional vane | Siemens Acvatix SQS65, 11/4" |
| 12 | Level transmitter | Foxboro, in/out: 0-6.6 kPa/4-20 mA |
| 13 | Power supply | Meanweal, NES-100-24 |
| 14 | System on/off switch | 16 A |
| 15 | Flow transmitter | ARF4 HH, turbine type |

A reservoir tank that is made of plastic in 40×30×30 cm dimensions at the bottom side of the set is utilized for the liquid reserve. A cylindrical tank that is made of transparent fiber-glass, which has a diameter of 15 cm, a height of 125 cm and a thickness of 4 mm, is employed for the control of the liquid level in the tank. The tope side of the tank is left open to the atmospheric pressure. The liquid level in the tank is sensed through Foxboro differential pressure transmitter. The transmitter has a measurement range of 0-6.6 kPa. The high pressure input of the transmitter $P_H$ is connected to the bottom side of the cylindrical tank ($P_H = P_{atm} + P_{liquid}$). Also, the low pressure input of the transmitter $P_L$ is left open to the atmospheric pressure ($P_L = P_{atm}$). Since both the top of the tank and the low pressure input of the transmitter are left open to the atmospheric pressure, the transmitter output changes depending on the height of the liquid level in the tank ($P_{difference} = P_H - P_L = P_{liquid}$). Water is used as the process variable to be controlled in the system. The specific gravity of water is about 998.2 kg/m3 at 1 atm and 20 °C. The laboratory temperature, in which there exists the experimental set, is about 20 °C. The specific gravity of water has stayed at about the same value during the experiments. In order to find out the base pressure of water in the cylindrical tank, the following equation is used

$$P = \rho.g.h .\qquad(5)$$

Where $P$ is the base pressure of the water (kPa), $\rho$ is the water density (kg/m3), $g$ is the acceleration of gravity (9.8 m/s2) and $h$ is the height of the water level (m). The height of the maximum water level calculated from (5) is about 0.7 m when considering that the transmitter measures a maximum pressure of 6.6 kPa (6,600 Pa). It means that the base pressure of the water is 0 kPa when the water level in the tank is equal to 0 cm, and this pressure is 6.6 kPa when the water level in the tank is equal to 70 cm. In the installation phase of the set, the output of the transmitter has been respectively adjusted 4 mA and 20 mA through a precision potentiometer on the transmitter when the tank is empty and full. Where the water density and the acceleration of gravity are constant and only variable is the water level. The Siemens Acvatix SQS65 proportional vane used for regulating the amount of the water from the reservoir is controlled by two different control output of 0-10 V DC, which belongs to the NC200 process control device and the SCADA control system with DAQ card. The vane completely closes the water input when 0 V DC is applied, and also completely opens the water input when 10 V DC is applied. The response time of the vane, the time required for passing from full open position to fully closed position or vice versa, is 25 s. The circulation of water is provided by a one phase AC induction motor of ½ HP manufactured by General Electric Company. In the system, the load change is manually realized by a flow control vane. In addition to the flow control vane, two manual flow adjustment vanes, which allow the input-output load changes in the system, are used to test the tuned controller parameters.

## 4. SOFTWARE

In the liquid level control system, the NC200 process control device that has been utilized as a compact PID controller has a DSP with 16 bits and contains on-off, P, PD, PI, PID controller algorithms as a standby code. There is a panel. User is able to tune the gain parameters connected with the control methods after selecting the controller type over the panel. Thus, the NC200 process control device has no need for additional software.

The software of the SCADA system with the DAQ card has been realized by the Advantech VisiDAQ software which is a Windows-based data acquisition, control, analysis and presentation development package. The feedback control algorithm with PID controller is executed on this software developed for the computer-based control. The block diagram of the developed software and preview image of the DAQ-SCADA window are given in **Fig. 4a**. In the diagram, while the level information from the A-4011L card is taken by the AI1 block, the output command is sent by the AO1 block to the proportional vane over the A-4021L card The data from the flow transmitter, which is only utilized in monitoring the flow rate of the liquid, is taken by the CTFQ1 block. NTCL1, SPIN1, SPIN2 and SPIN3 tags are used to enter the values of set point, Kp, Ki and Kd, respectively. The PID1 block includes the PID control algorithm which is required to obtain the control signals. The output position of the system is regulated by PRG2 and PRG3 user programs. Preview image of the DAQ-SCADA window which appears on the computer screen when running the software is shown in **Fig. 4b**.

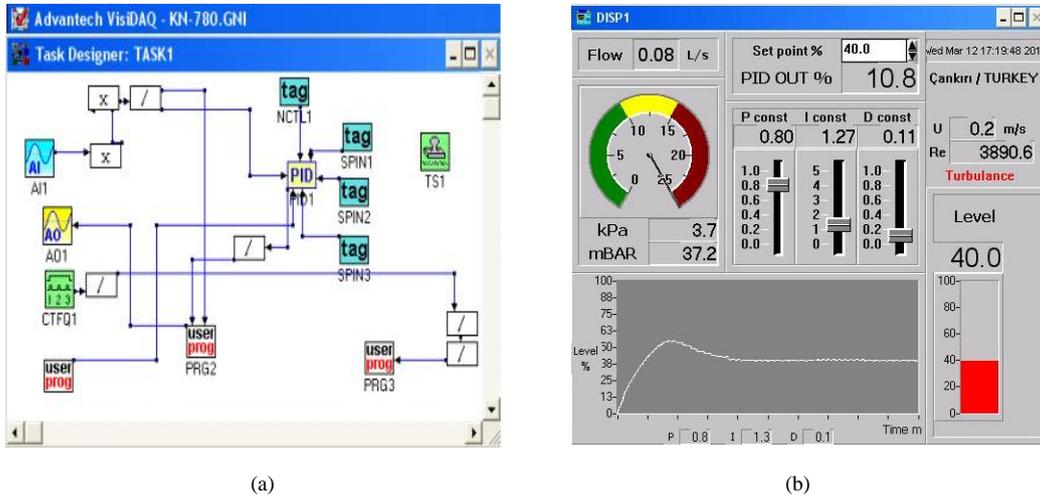

(a) (b)

**Figure 4.** (a) The block diagram of the DAQ-SCADA software and (b) preview image of the DAQ-SCADA window

## 5. RESULTS AND DISCUSSION

In this section, we are going to separately present the experimental results and evaluations of the liquid level control training set which has been controlled by both the NC200 process control device and the computer-based SCADA system with the DAQ card.

### 5.1. Experimental results and evaluations of the training set controlled by the NC200 process control device

As the compact controller, the control of the liquid level control system has been carried out by using on-off, P, PD, PI and PID controllers over the NC200 process control device.

For the system controlled by the on-off controller, the process variable and the controller hysteresis range have been set the values of 70% (height of 49 cm) and 10% (7 cm), respectively. When the level value exceeds 49 cm, the controller completely closes the inflow liquid into the tank by sending 0 V DC to the proportional vane. On the other hand, when the level value of the tank drops below 42 cm (49 cm – 7 cm = 42 cm), the controller entirely opens the inflow liquid into the tank by sending 10 V DC to the proportional vane and allows filling the tank with liquid. Besides these values, we also have tested the system for the different values of set point and hysteresis range.

Once the flow adjustment vane, which is manually controlled and regulates the inflow liquid into the tank, is turned the highest value, the liquid level reaches 8 cm above of the set value (49 cm + 8 cm = 58 cm). This is due to the response time which occurs when the proportional vane passes from full open position to fully closed position. This problem in the on-off control can be tackled with using a solenoid vane instead of a proportional vane. However, solenoid vanes are not used in the systems controlled by PID controllers since these controllers adjust the output variable to the set value by producing an output proportional to the error. Therefore, a proportional vane has been used in the designed liquid level control training set.

When the liquid level of the operating process is dropped to 42 cm, which is the lower limit of the hysteresis range, a decrease has been seen up to about 8 cm (42 cm – 8 cm = 34 cm). This situation has been also tested for the adjusted different set point and hysteresis values.

In order to control the system with P, PD, PI and PID controller methods, according to the Z-N method, firstly, $K_p$, $K_i$ and $K_d$ values have been set 1, 9999 and 0, respectively. The NC200 process control device has been taken the PID mode. Then, the process variable has been set 70%. The controller gain parameters in the device have been tuned by the Z-N method. Then, the training set has been operated with the adjusted set point value and $K_p$ has been decreased by small increments until a sustained oscillation with fixed amplitude is obtained. This oscillation is reached when $K_p$ is 80 and the gain value has been saved as Ku. The oscillation period $P_u$ at this moment has been measured as approximately 36 s. The measurement procedure of the oscillation period is based on the average of the fifty periods counted by a chronometer. The highest value and lowest value of these periods have been obtained as 39 s and 34 s, respectively. The differences among the period values have arisen from the reading errors and the changes of the process variable.

After determining $K_u$ and $P_u$ values, the final values of $K_p$, $K_i$ and $K_d$ for each controller have been calculated employing the equations in Ziegler and Nichols [33], as shown in **Table 3**. The experimental results of the system, which is controlled by the P, PD, PI and PID controllers with the calculated gain parameters, are shown in **Fig. 5**.

**Table 3.** The Controller gain parameters tuned by the Z-N Method for the NC200 process control device ($K_u$ = 80, $P_u$ = 36 s)

| Control type | $K_p$ | $K_i$ | $K_d$ |
|---|---|---|---|
| P | 40 | - | - |
| PD | 36 | - | 162 |
| PI | 36 | ~1* | - |
| PID | 48 | ~3* | 216 |

*: 1.20 and 2.66 values could not be entered to the controller. Instead of them, 1 and 3 values have been entered.

The output response curve of the P controlled system is given in **Fig. 5a**. For $K_p$ = 40 value of the P controller, the system has come to the steady state at 72 s and the steady state error has been obtained as 8.16%.

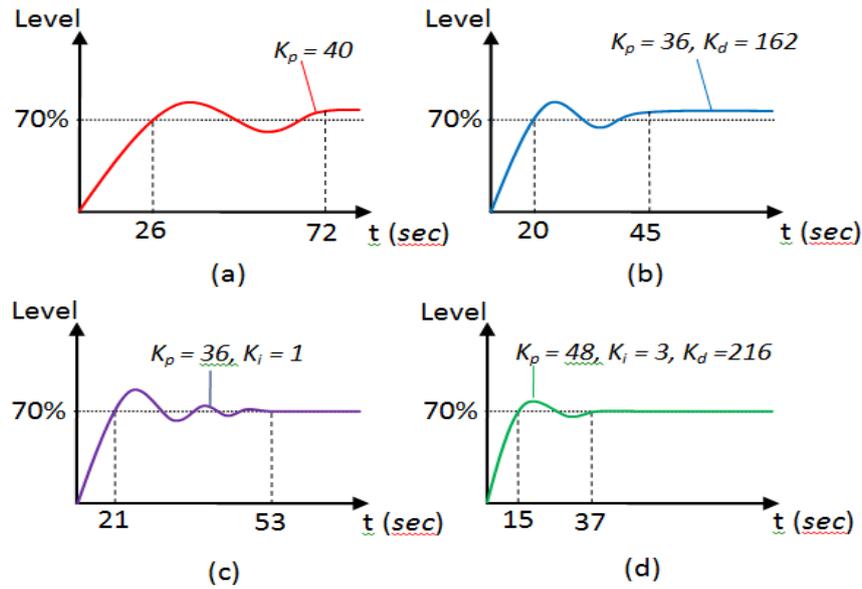

**Figure 5.** Experimental results of the liquid level system controlled by the NC200 process control device (a) P, (b) PD, (c) PI and (d) PID.

In **Fig. 5b**, the output response curve of the PD controlled system is shown. For $K_p = 36$ and $K_d = 162$ values of the PD controller, although the system has reached the steady state at 42 s, the steady state error has not been eliminated again. The steady state error has been 10.2% for the mentioned gain parameters. As understood from the obtained results, the adding to the P control structure of D control has only allowed the system to come to the steady state more quickly.

In **Fig. 5c**, the output response curve of the PI controlled system is indicated. The operating system with $K_p = 36$ and $K_i = 1$ has reached the steady state at 52 s. and the steady state error has been zeroized. The deviations that occur until reaching the set value have been fairly reduced by the PI control method, when comparing to the P and PD control methods.

In **Fig. 5d**, the output response curve of the system controlled by the PID controller that is widely used in the systems with slow response time is shown. When the system is operated with the gain values calculated for the PID controller in **Table 3**, the settling time and the steady state error has been obtained as 37 s. and 0%, respectively. The deviations preceding the steady state of the system have been much less than the other three control methods.

In addition to these analyses, the liquid level control system has been also tested by different gain values of the controllers, but it is observed that the system has had better output response with the gain parameter values.

### 5.2. Experimental results and evaluations of the training set controlled by the computer-based SCADA system

A SCADA system with DAQ card developed by authors has been used in control of the system, as a computer-based controller. In the system, the PID controller gain parameters has been determined by fine tuning to the gain parameters, which have been previously obtained by the Z-N method, on the SCADA screen given in **Fig. 4b**. According to the mentioned procedure, the tuned optimum controller gains have been obtained as $K_p = 0.80$, Ki = 1.27 and $K_d = 0.11$ for the set value of 40% of the process variable. On the SCADA screen, the chart at the bottom of the window shows the change of the process variable versus time and this change is also monitored by a bar graph which is located at the right side of the chart. The vertical and horizontal axes on the chart are scaled as percent (%) and minute, respectively. The maximum deviation of the process variable from the set value, which occurs before the system comes to the steady state, has been measured as 15% in negative direction. The system has reached the steady state at 4 min and the steady state error has been obtained as 0%. The transient response to the parameter changes of the system has been tested over the set point, the output load and the input load. **Fig. 6** shows the transient responses to these changes of the liquid level control system which is controlled by the computer-based SCADA system.

**Fig. 6a, b,** and **c** depict the transient response curves of the system for the case when the set point is changed from the current value to the higher value. The changes in the set point has been fulfilled for three different cases; namely from 40% to 50%, from 50% to 60% and from 60% to 80%. The transient responses to these changes of the system have been obtained as about 1 min, 50 s, and 4 min, respectively. The steady state errors measured for three cases have been 0 %. When the set value is dropped from 80% to 45%, a deviation of 3% has occurred in the output response of the system in accordance with the new set value, as understood from the **Fig. 6d**. The system has corrected this deviation at about 2.40 min and zeroized the steady state error.

In **Fig. 6e**, the transient response of the system which operates in the set value of 45% is given for the case when the output load is decreased by the manual vane. In that case, the deviation of 2% from the set value in the process variable has been corrected by the transient response of about 2 min and the steady state error has been zeroized.

For the case that the output load is abruptly dropped while the system is working with the last steady state conditions in **Fig. 6e**, the measured transient response is illustrated in **Fig. 6f**. In that case, a deviation of 23% has happened in the process variable and this has been corrected at about 5 min. The steady state error has been obtained as 0%.

In **Fig. 6g**, the transient response is shown for the case that the input load is increased the highest value by the manual controlled flow adjustment vane while the system is operating without changing the set value and the output load. In that case, a deviation of 5% from the set point has been taken place in the process variable and this has been corrected at about 2 min. However, the steady state error has been obtained as 2%.

In **Fig. 6h**, the transient response is given for the case that the set value of the system operating with the last steady state conditions in **Fig. 6g** is changed from 45% to 60%. The deviation arisen from the change has been corrected at about 40 s. The steady state error has been occurred as 2%, as in the previous case.

In **Fig. 6i**, the transient response is shown for the case that the set value of the system is changed from 60% to 70% while the input load is being kept the highest value. In that case, the process variable has settled to the set value at about 50 s and the steady state error has stayed like the previous one.

In **Fig. 6j**, for the case that the input load of the system working with the last steady state conditions in **Fig. 6i** is manually decreased by the flow adjustment vane, the transient response is given. In that case, the transient response of the system to the deviation of 10% occurring in the process variable has been about 2 min and the steady state error has been zeroized.

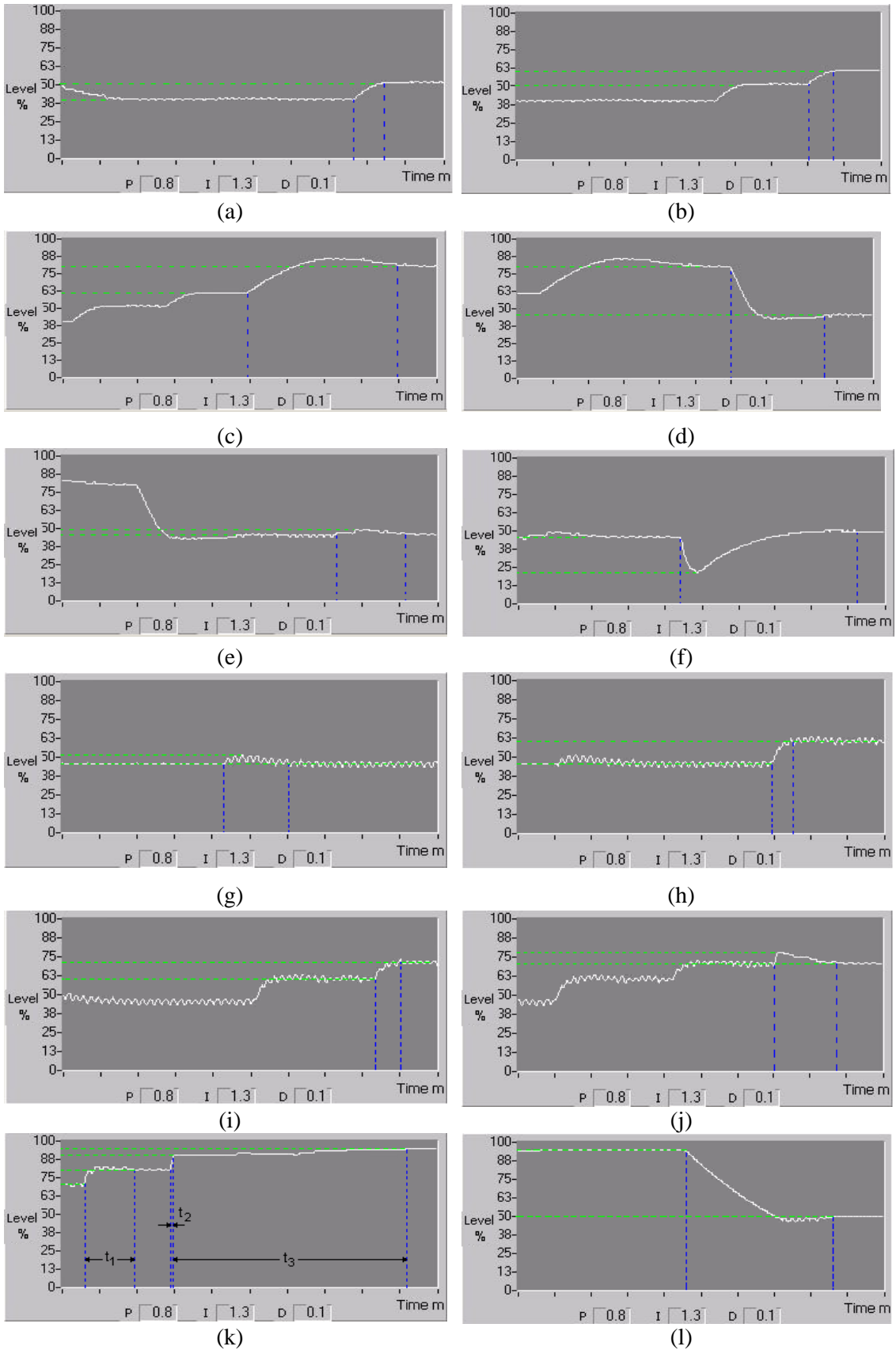

**Figure 6.** Transient responses of the liquid level control system controlled by the computer-based SCADA system for $K_p = 0.80$, $K_i = 1.27$ and $K_d = 0.11$

In **Fig. 6k**, the transient response is illustrated for the cases that the set value of the system is changed from 70% to 80%, from 80% to 90% and from 90% to 95%, consecutively. In the case of the change from 70% to 80% of the set value, the transient response of the system is obtained as about $t_1$ = 1.20 min and the steady state error has been 2%. When the set value is changed from 80% to 90%, the process variable has come the steady state in a very short time of about $t_2$ = 5 s along with the zero steady state error. The system has reached the steady state in a long time about $t_3$ = 6.20 min along with the zero steady state error when the set value is changed from 90% to 95%.

The last curve in **Fig. 6l** shows the transient response of the system for the case that the set point is dropped by a large change from 95% to 50%. In that case, the steady state error has been eliminated and the deviation of 3% resulted from the change has been corrected at about 4 min.

As a result, the fundamental control methods commonly used in automatic control systems could be practically and easily applied on the training set. The system analyzes could be realized in detail over the DAQ-SCADA interface, on which the parameters related to the system and controller can be easily tuned and the experimental results can be graphically displayed. Furthermore, the presence of a compact PID controller device in the training set would provide significant advantages in terms of experiment interruptions both when the comparative analyzes related to the system are carried out and when a problem in the PC is happened.

## 6. FEEDBACK FROM STUDENTS

The assessment of the designed liquid level control training set developed for engineering education courses has been conducted on seventeen students who take the automatic control system course. Firstly, basic control methods have been described to the students and then fundamental knowledge's have been provided about system controls. After that, the mathematical model of the liquid level control system has been derived and the simulation studies have been performed in MATLAB/Simulink environment by using this model. The application study of the liquid level control system has been started by the training set after the theory and simulation related to the system were completed.

In the first session of the application study, the NC200 process control device has been activated and then the effects on the performance of the liquid level control system of on-off, P, PD, PI and PID controllers have been experimentally investigated and analyzed for the cases where both the system was worked under different conditions and the gain parameters of the controllers were changed. In the second session of the application study, the performance analyses of the liquid level control system controlled by the computer-based SCADA system have been realized under the same working conditions and parameter changes.

At the end of this training, a form that contains the following six questions has been directed to the sample student group:

    **a.** Before studying with the liquid level control training set, the theory of the fundamental feedback control methods in the automatic control systems, including on-off, P, PD, PI and PID controllers, has been understood.

    **b.** After learning the theory of the controllers, the realization of their applications has greatly enhanced the understandability of the subject.

    **c.** The main differences between the compact PID controller and the computer-based SCADA system have been clearly comprehended.

d. If I encounter with a similar process in industry, then I can easily adjust the controller gain parameters.

e. In the computer-based control, the having a flexible and visual screen of the DAQ-SCADA system has provided us an opportunity to comprehend and evaluate the experimental results in a good way.

After the theory, the practices based on both the compact PID control and the DAQ-SCADA systems are unnecessary and it is sufficient to apply only one of the control structures.

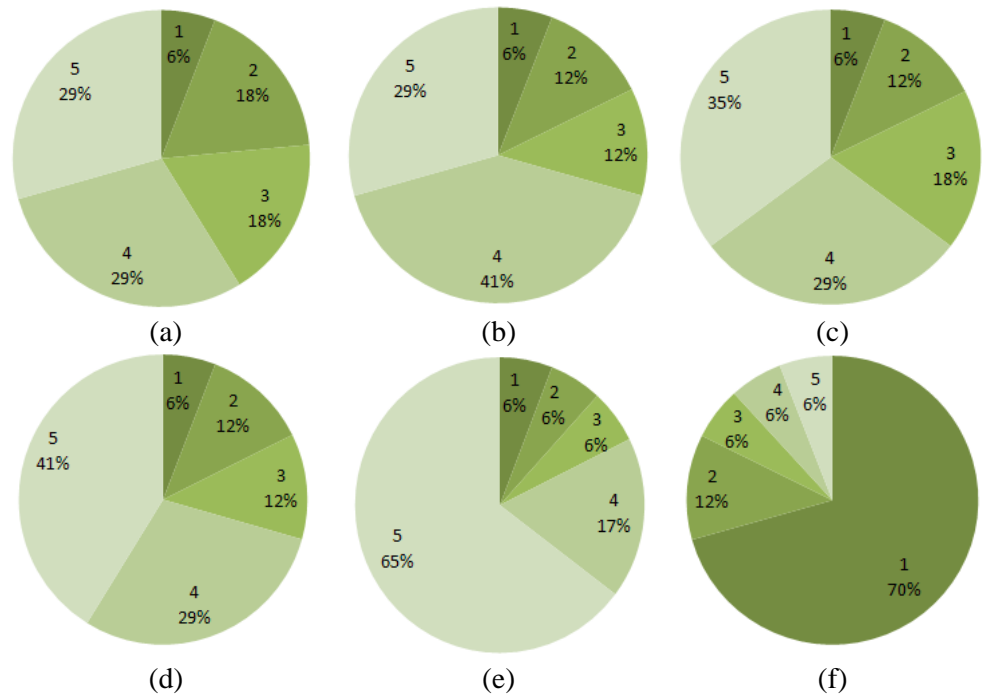

**Figure 7**. The survey on the automatic control students for the liquid level control training set.

The students have been requested to mark one of the following answers according to the degree of agreement with the statement: (1) "*strongly disagree*", (2) "*disagree*", (3) "*neither agree nor disagree*", (4) "*agree*" and (5) "*strongly agree*". The received feedbacks from the six questions are given in detail in **Fig. 7**. The survey on the automatic control students has shown that the liquid level control training set would be both increase the learning ability of the students and allow them to express to be confidently worked in industry.

## 7. CONCLUSION

In this study, a liquid level control training set, on which the students could separately both theoretically and experimentally apply the feedback control algorithms, have been designed for control engineering courses. The practical application of this set realized on a prototype by both a computer-based control and a DSP-based control. The results of the experimental studies performed on the training set clearly shown the following acquirements: (1) the fundamental control algorithms described in the control engineering courses could be successfully applied to the control of a process, (2) thanks to the newly developed DAQ-SCADA software containing a visual and flexible interface, a process could be controlled by the computer-based control structures and analyzed under the different operating conditions in detail, (3) the comparative analyses between the computer-based control and the DSP-based control structures could be realized. It can be concluded from the feedbacks received from the students that the training set provides an important contribution to the understanding and evaluating of the algorithm and controller structures used in control of a process. With these features, the developed set is a useful and beneficial instrument for the control engineering courses.

## Authors

**Hayati Mamur** received the B.S. degree from Department of Electrical Education, Gazi University, in 1996, has graduated the M.S. and Ph.D. degrees from Department of Electronics and Computer Education, Gazi University, Ankara, Turkey, in 2005 and 2013, respectively. He is working as an Asst. Prof. at the Department of Electrical and Electronics Engineering, Faculty of Engineering, Manisa Celal Bayar University, Manisa, Turkey. His current research interests include micro wind turbines, permanent magnet generators, convertors, invertors, micro grid systems and thermoelectric modules.

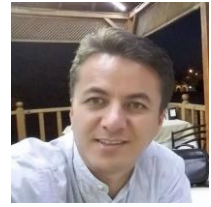

**Ismail Atacak** received the B.S., M.S. and Ph.D. degrees from Department of Electronics and Computer Education, Gazi University, in 1994, 1998, and 2005, respectively. From 2007 to 2012, he worked as an Assistance Professor at the Department of Electronics and Computer Education, Faculty of Technical Education, Ankara, Turkey. He is currently working as an Assistance Professor at the Department of Computer Engineering, Faculty of Engineering, Ankara, Turkey. His research interests include power systems, artificial intelligent based algorithms, optimization based algorithms and Engineering Education.

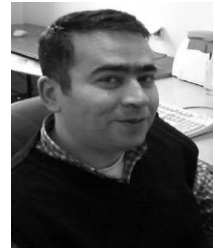

**Fatih Korkmaz** was born in Kırıkkale, Turkey, in 1977. He received the B.S., M.S., and Ph.D. degrees in in electrical education, from University of Gazi, Ankara, Turkey, respectively in 2000, 2004 and 2011. Since 2012, he is working as an Asst. Prof. Dr. at the Department of Electrical and Electronics Engineering, Faculty of Engineering, Cankiri Karatekin University, Cankiri, Turkey. His current research field includes Electric Machines Drives and Control Systems.

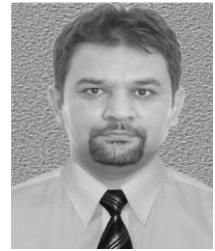

**Mohammad Ruhul Amin Bhuiyan** received the B.Sc., M.Sc. degrees in Applied Physics and Electronic Engineering from Rajshahi University, Bangladesh in 1994 and 1995, respectively. Ph.D. degree in Applied Physics, Electronics and Communication Engineering from Islamic University, Bangladesh in 2008. Now at present he is working as a visiting scientist at the Department of Electrical and Electronics Engineering, Faculty of Engineering, Manisa Celal Bayar University, Manisa, Turkey.

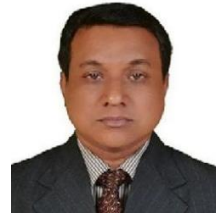